\documentclass[journal]{IEEEtran}
\usepackage{cite}
\usepackage{comment}
\usepackage{booktabs}
\usepackage{algpseudocode}
\usepackage{graphicx}
\usepackage{mathtools}
\usepackage{amsmath}
\usepackage{amssymb}
\usepackage{bm}
\usepackage{subfig}
\usepackage{stfloats}
\usepackage{multirow}
\usepackage{graphicx}
\usepackage{textcomp}
\usepackage{caption}
\usepackage{float,lscape}
\usepackage{longtable}
\usepackage{caption}
\usepackage{url}
\usepackage{tabulary}
\usepackage{lipsum, color}
\usepackage{enumitem}
\usepackage{mathtools, cuted}

\usepackage{amsmath,tikz}
\usepackage{amssymb}
\usepackage{multirow}
\usepackage{array}
\usepackage{enumerate}
\usepackage{calligra}
\usepackage[linesnumbered,ruled] {algorithm2e}
\usepackage{lipsum, color}
\usepackage[bottom]{footmisc}

\ifCLASSINFOpdf
 
\else
 
\fi

\begin{document}
\title{Quantum-Driven State-Reduction for Reliable UAV Trajectory Optimization in Low-Altitude Networks}
\author{Zeeshan Kaleem,~\IEEEmembership{Senior Member,~IEEE}, Muhammad Afaq, Chau Yuen,~\IEEEmembership{Fellow,~IEEE}, \\Octavia A. Dobre,~\IEEEmembership{Fellow,~IEEE}, John M. Cioffi,~\IEEEmembership{Life Fellow,~IEEE}
\thanks{This work is supported by KFUPM Ibn Battuta Global Scholarship grant with number ISP24202.
Zeeshan Kaleem and Muhammad Afaq are with the Department of Computer Engineering and with IRC for Smart Mobility and Logistics, and IRC Intelligent and Secure Systems, King Fahd University of Petroleum \& Minerals (KFUPM), Dhahran 31261, Saudi Arabia (e-mail: zeeshankaleem@gmail.com, muhammad.afaq@kfupm.edu.sa)

Chau Yuen is with the School of Electrical and Electronics Engineering, Nanyang Technological University, Singapore (email: chau.yuen@ntu.edu.sg).

O. A. Dobre is with the Faculty of Engineering and Applied Science,
Memorial University, Canada (email: odobre@mun.ca)

John M. Cioffi is with the Department of Electrical Engineering, Stanford University, Stanford, CA 94305 USA (e-mail: cioffi@stanford.edu).

}
\vspace{-0.35in}}

\maketitle
\vspace{-0.19in}
\begin{abstract}
This letter introduces a Graph-Condensed Quantum-Inspired Placement (GC-QAP) framework for reliability-driven trajectory optimization in Uncrewed Aerial Vehicle (UAV)–assisted low-altitude wireless networks. The dense waypoint graph is condensed using probabilistic quantum-annealing to preserve interference-aware centroids while reducing the control state space and maintaining link-quality. The resulting problem is formulated as a priority-aware Markov decision process and solved using $\varepsilon$-greedy off-policy Q-learning, considering UAV kinematic and flight corridor constraints. Unlike complex continuous-action reinforcement learning approaches, GC-QAP achieves stable convergence and low outage with substantially and lower computational cost compared to baseline schemes.
\end{abstract}
\IEEEpeerreviewmaketitle
\vspace{-0.1in}
\section{Introduction}
The low-altitude wireless networks employs Uncrewed aerial vehicle (uav) as aerial base stations to enable flexible and resilient connectivity. However, multi-UAV operation introduces interference, energy, and outage challenges, necessitating efficient placement and trajectory optimization for reliable, low-latency service~\cite{zhao2024quantum}, \cite{pan2023joint}. 
The increasing severity of disasters necessitates rapid-deployment communication architectures. UAV-enabled low-altitude wireless networks provide a flexible solution, but large state-space optimization challenges motivate quantum computing (qc) for scalable and efficient UAV placement in critical scenarios~\cite{kaleem2025quantum},\cite{10318071}. 

Existing UAV placement methods often rely on complex optimization that converges slowly and struggles with real-time adaptability. For example, \cite{10539386} jointly optimizes phases, scheduling, and UAV trajectory via alternating optimization; it converges quickly per run but incurs high runtime and poor scalability due to large state space and costly per-iteration updates. Likewise, \cite{liu2022fair} addressed fair coverage and energy efficiency through constrained optimization, yet incurred high computational cost across time-varying scenarios.
Reinforcement learning (rl) has been applied to optimize UAV-user association, achieving improved sum-rate and energy efficiency \cite{kaleem2024reinforcement}. The authors in \cite{ibanez2025optimizing} adopted proximal-policy optimization (ppo) with continuous actions to optimize UAV flights for emergency connectivity, focusing on throughput and coverage. However, it lacks outage-based objectives, power control, and state-space reduction, requiring intensive training and computation. Although these methods report performance improvement, they rely on dense airspace, leading to scalability issues as state-action spaces grow with waypoints and interference coupling. This increases training time and energy consumption, limiting their practicality in disaster scenarios.
QC is emerging as a powerful alternative to classical methods, offering improved time, energy, and reliability performance for future communication systems \cite{zhao2024quantum}. In UAV networks, quantum-inspired techniques have been applied to path planning. For example, \cite{li2022path} proposed a DRL navigation framework with quantum-inspired experience replay, which enhanced decision-making and reduced outage, outperforming conventional DRL and non-learning baselines. Recent works have explored quantum techniques for UAV mission planning and trajectory optimization. Near-term quantum algorithms and hybrid quantum–classical frameworks have shown speed improvements for routing, clustering, and resource allocation \cite{Jeong2025QA_MultiUAV}. 
Layerwise Quantum-based DRL (lq-drl) has been proposed to handle large, continuous UAV optimization problems \cite{10153404}, while quantum-inspired RL (qiRL) leverages collapse probability and amplitude amplification for efficient trajectory planning in uplink scenarios \cite{9456900}.
Although existing studies have attempted to optimize UAV path planning by reducing the state space, they still overlook key challenges such as inter-UAV interference and the persistence of large search spaces. As a result, the curse of dimensionality remains the principal bottleneck in multi-UAV trajectory optimization. Motivated by this, we propose a graph–condensation (GC) strategy via quantum-annealing for placement (GC-QAP) optimization. The proposed GC-QAP is interference-aware multi-UAV scheme for trajectory optimization. The GC-QAP uses a quantum-inspired graph condensation framework that compresses the dense UAV waypoint set using probabilistic annealing, retaining high-value states while drastically shrinking the search space. This condensation produces task-faithful subgraphs that preserve coverage and interference structure with minimal performance loss and substantial computational savings. Second, GC-QAP incorporates signal-to-interference and noise ratio (sinr)-based user association with uplink power control to ensure realistic interference modeling. Third, GC-QAP trains coordinated multi-UAV policies through independent learning to minimize outage for priority users. Overall, the proposed GC-QAP framework achieves significantly lower outage and faster convergence compared to existing baselines.
\vspace{-0.28in}
\section{System Model}
This model is an uplink network assisted by $N$ UAV aerial base station (ABS) that serves $K$ single–antenna ground users (GUs) over bandwidth $B$ (Hz). The terrestrial service region consists of
$\mathcal{A}=[x_{\min},x_{\max}]\times[y_{\min},y_{\max}]$.
ABS $n\!\in\!\mathcal{N}\triangleq\{1,\dots,N\}$ flies at fixed altitude $h_n$ and horizontal position
$\mathbf{q}_n[t] = [x_n[t],y_n[t]]^\top$ at discrete slot $t=1,\dots,T$.
Its 3D location is $\mathbf{u}_n[t]=[x_n[t],y_n[t],h_n]^\top$. Terrestrial user $k\!\in\!\mathcal{K}\triangleq\{1,\dots,K\}$ is located at
$\mathbf{w}_k=[x_k,y_k,0]^\top$.
A subset $\mathcal{K}^{\sf pr}\subset\mathcal{K}$ are \emph{priority} users; $\mathcal{K}^{\sf nr}\!\!=\!\mathcal{K}\setminus\mathcal{K}^{\sf pr}$ are non-priority users. In each slot, users transmit simultaneously on shared UL resources to generate interference-limited regime. Trajectory updates occur per slot subject to motion constraints. 

To reduce the ABS' continuous placement space, the model precomputes a set of $M$ candidate waypoints (centroids) represented as $\mathcal{C}=\{\mathbf{c}_m\in\mathbb{R}^2\}_{m=1}^M$ by using schemes like QA and k-means. Therefore, the $n$-th ABS can occupy only these waypoints $\mathbf{q}_n[t]\in\mathcal{C},\quad \forall n,t.$
Feasible moves are encoded by an undirected graph $\mathcal{G}=(\mathcal{C},\mathcal{E})$ with adjacency matrix $\mathbf{A}$; a transition $\mathbf{q}_n[t]\!\to\!\mathbf{q}_n[t\!+\!1]$ is allowed only if $(\mathbf{q}_n[t],\mathbf{q}_n[t+1])\in\mathcal{E}$. A per-slot step constraint enforces bounded horizontal speed, computed as $\|\mathbf{q}_n[t\!+\!1]-\mathbf{q}_n[t]\|_2 \le v_{\max}\Delta_t,\ \ \forall n,t. $ 
Let $d_{k,n}[t]=\|\mathbf{u}_n[t]-\mathbf{w}_k\|_2$ be the 3D distance and 
$\theta_{k,n}[t]=\arcsin\!\big(\tfrac{h_n}{d_{k,n}[t]}\big)$ the elevation angle in radians, with $\theta^\circ_{k,n}[t]=\tfrac{180}{\pi}\theta_{k,n}[t]$ in degrees. The line-of-sight (los) probability follows the model $P^{\sf LoS}_{k,n}[t] = \big[b_1\left(\theta^\circ_{k,n}[t]-\xi\right)\big]^{b_2}, \quad $, and therefore, Non-LoS (nlos) can be computed as $P^{\sf NLoS}_{k,n}[t] = 1 - P^{\sf LoS}_{k,n}[t],$ where it is between $[0,1]$. The large-scale loss (dB) combines free-space reference $K_0$ and distance exponent $\alpha$ with LoS/NLoS excess losses $\kappa_{\sf LoS}$, $\kappa_{\sf NLoS}$, which can be computed as
$L_{k,n}^{\sf eff}[t]\,[\mathrm{dB}] = 10\log_{10}K_0 + 10\alpha\log_{10}d_{k,n}[t]
+ 10\log_{10}\!\big(P^{\sf LoS}_{k,n}\kappa_{\sf LoS} + P^{\sf NLoS}_{k,n}\kappa_{\sf NLoS}\big).$ The small-scale fading is modeled as i.i.d. and assume $|f_{k,n}[t]|^2\sim\exp(1)$. Hence, the total linear channel gain is $ g_{k,n}[t] = 10^{-L_{k,n}^{\sf eff}[t]/10}\,|f_{k,n}[t]|^2.$ To reduce the interference, each user applies open-loop power control per slot, and the power can be computed as $P^{\sf tx}_k[t]\,[\mathrm{dBm}] = \min\!\big\{P_{\max},\; P_0 + \alpha_{\sf OL}\,PL_{k,\hat{n}_k[t]}[t] + 10\log_{10}RB\big\},$ where $P_{\max}$ is the user maximum power, $P_0$ the target baseline, $\alpha_{\sf OL}\in[0,1]$ the compensation factor, $RB$ the number of resource blocks (rb) allocated, and $PL_{k,\hat{n}}$ the path loss (dB) with respect to the currently associated ABS $\hat{n}_k[t]$. 
Given the instantaneous received power, the received power is $P^{\sf rx}_{k,n}[t] = p^{\sf tx}_k[t]\; g_{k,n}[t],$, where $p^{\sf tx}_k[t]$ is the power in watts. Each user associates to the ABS providing the maximum received power, that is, $\hat{n}_k[t] \in \arg\max_{n\in\mathcal{N}} \; P^{\sf rx}_{k,n}[t].$

Let $\mathcal{U}_n[t] = \{k : \hat{n}_k[t] = n\}$ denote the set of users served by the $n$-th ABS in slot $t$. 
Assuming full-frequency reuse, the uplink interference observed at the $n$-th ABS for user $i \in \mathcal{U}_n[t]$ 
arises from users served by other ABSs, and can be computed as $I_{n,i}[t] = \sum_{\substack{j\in\mathcal{K}\setminus\{i\}}} \mathbf{1}\{\hat{n}_j[t]\neq n\}\; p^{\sf tx}_j[t]\; g_{j,n}[t]$, where $I_{n,i}[t]$ is the interference at ABS $n$ for user $i$ at time $t$, $\hat{n}_j[t]$ is the ABS serving user $j$ at time $t$, $\mathbf{1}\{\cdot\}$ is the indicator function.
The instantaneous SINR $\gamma_{i}[t] $ of user $i$ received at its serving ABS $n=\hat{n}_i[t]$ is $
\gamma_{i}[t] = \frac{p^{\sf tx}_i[t]\; g_{i,n}[t]}{\sigma^2 + I_{n,i}[t]},$
where, $\sigma^2$ is the thermal noise power in Watts. 
Data rate achieved by each user is computed as $
R_i[t] = B \log_2\!\big(1+\gamma_{i}[t]\big).$
A user is considered in outage if $\gamma_{i}[t] < \gamma_{\sf th}$, where $\gamma_{\sf th}$ is the threshold SINR. 
Note that the probability of outage for user $i$ at slot $t$ can be expressed as $\mathbb{P}\!\left(\gamma_i[t] < \gamma_{\sf th}\right) 
= \mathbb{E}\!\left[\mathbf{1}\{\gamma_i[t] < \gamma_{\sf th}\}\right].$

Accordingly, the per-slot outage of the $n$-th ABS and the overall network outage are expressed as $P_{\text{out},n}[t] = \frac{1}{|\mathcal{U}_n[t]|}\sum_{i\in\mathcal{U}_n[t]} 
\mathbb{P}\!\left(\gamma_{i}[t]<\gamma_{\sf th}\right)$ and $P_{\text{out,net}}[t] = \frac{1}{K}\sum_{i=1}^K 
\mathbb{P}\!\left(\gamma_{i}[t]<\gamma_{\sf th}\right)$, respectively.
To capture service differentiation, we further define the 
\emph{priority-based outage probabilities} as
\begin{equation}
P_{\text{out}}^{(c)}[t] 
= \frac{1}{|\mathcal{K}^{(c)}|}
\sum_{i\in\mathcal{K}^{(c)}} 
\mathbb{P}\!\left(\gamma_i[t] < \gamma_{\sf th}\right),
\quad c \in \{\sf pr, nr\},
\end{equation}

\section{Problem Formulation}
The objective formulates the UAV trajectory optimization as an outage-aware reliability maximization problem, where UAV trajectories are optimized to minimize outage probabilities, particularly for priority users. The optimization problem is
\begin{subequations}\label{obj}
\begin{align}
\min_{\{\mathbf{q}_n[t]\},\,\{p^{\sf tx}_i[t]\}} \quad
& \sum_{t=1}^T \Big( 
\mu_{\sf pr}\, P_{\text{out}}^{(\sf pr)}[t] 
+ \mu_{\sf nr}\, P_{\text{out}}^{(\sf nr)}[t]
\Big), \label{obj:outage}\\
\text{s.t.}\quad
& (x_n[t],y_n[t]) \in \mathcal{C}, \quad \forall n,t \label{c1}\\
& (x_n[t{+}1],y_n[t{+}1]) \in \mathcal{N}\!\big((x_n[t],y_n[t])\big), \quad \forall n,t \label{C2}\\
& \|\mathbf{q}_n[t{+}1]-\mathbf{q}_n[t]\|_2 \le v_{\max}\Delta_t, \quad \forall n,t \label{C3}\\
& h_{\min} \le h_n[t] \le h_{\max}, \quad \forall n,t \label{C4}\\
& 0 \le p^{\sf tx}_i[t] \le P_{\max}, \quad \forall i,t \label{C5}
\end{align}
\end{subequations}
where $\mu_{\sf pr},\mu_{\sf nr}\in[0,1]$ are outage weights with typically $\mu_{\sf pr}\!\gg\!\mu_{\sf nr}$ to protect priority users.
The constraint \eqref{c1} restricts each UAV to feasible
positions on the condensed waypoint graph $\mathcal{C}$, while \eqref{C2} enforces that UAV motion occurs only along adjacent nodes. The kinematic constraint
\eqref{C3} limits the maximum horizontal displacement of a UAV in a
time slot, and \eqref{C4} limits its altitude within the range. Finally, the power constraint \eqref{C5} ensures that the user
transmit power remains within the allowed budget.

\subsection{Markov Decision Process (MDP) Representation of Problem}
The UAV trajectory optimization is a MDP and so enables learning through RL. 
At each time slot $t$, the system state is $\mathbf{s}[t] = \{\mathbf{q}_1[t],\dots,\mathbf{q}_N[t],\, \gamma_1[t],\dots,\gamma_K[t]\}$,
where $\mathbf{q}_n[t]$ denotes the position of UAV~$n$ and $\gamma_i[t]$ is the instantaneous SINR of user~$i$. 
The action $\mathbf{a}[t]$ corresponds to the next waypoint selection for each UAV from the condensed waypoint graph $\mathcal{C}$, 
while the environment transition follows the UAV motion and power constraints~\eqref{c1}--\eqref{C5}. 
The RL agent learns a policy $\pi(\mathbf{a}[t] \mid \mathbf{s}[t])$ that minimizes outage and maximizes reliability over time.
UAV trajectories training designs a \emph{priority-aware reward function} that directly reflects per-user reliability and UAV-level outage penalties. 
The instantaneous reward for UAV~$n$ at time~$t$ is
\begin{align}
r_n[t] &\triangleq
-\mu_{\sf pr}\!\!\!\!\sum_{i \in \mathcal{U}_n^{\sf pr}[t]}\!\!\!\!
\mathbf{1}\{\gamma_i[t] < \gamma_{\sf th}\}
-\mu_{\sf nr}\!\!\!\!\sum_{i \in \mathcal{U}_n^{\sf nr}[t]}\!\!\!\!
\mathbf{1}\{\gamma_i[t] < \gamma_{\sf th}\}  \nonumber\\
&\quad
-\mu_{\sf pr}\, \phi^{\sf pr}_n[t]
-\mu_{\sf nr}\, \phi^{\sf nr}_n[t],
\label{eq:reward2}
\end{align}
where $\phi^{\sf pr}_n[t]$ and $\phi^{\sf nr}_n[t]$ denote the outage fractions of priority and non-priority users served by UAV~$n$, respectively.
The total system reward is $\mathcal{R}[t] = \sum_{n=1}^N r_n[t]$, and the RL agent seeks to maximize the expected discounted return, $\max_{\pi}\;\; \mathbb{E}_{\pi}\!\left[\sum_{t=1}^T \zeta^{t-1} \mathcal{R}[t]\right],$ where $\zeta \in (0,1)$ is the discount factor.

The proposed reward formulation in \eqref{eq:reward2} is consistent with the optimization objective in \eqref{obj}, where weighted normalized rates encourage UAV trajectories that improve coverage and throughput, and outages are explicitly penalized with stronger penalties $\mu_{\sf pr}$ for priority users, 
and UAVs trajectories are jointly optimized to reduce the outage probability. By considering this reward in the MDP framework, RL agents learn the trajectory and association policies that converge towards the solution of \eqref{obj}.

For each UAV $n\!\in\!\{1,\dots,N\}$, GC-QAP learns an independent state–action value
$Q_n:\mathcal{C}\times\mathcal{A}\!\to\!\mathbb{R}$ over the condensed waypoint set $\mathcal{C}$.
At time $t$, UAV $n$ observes state $s_n[t]\!\in\!\mathcal{C}$, selects an action
$a_n[t]\!\in\!\mathcal{A}(s_n[t])\!\triangleq\!\mathcal{N}(s_n[t])$, moves to $s_n'[t]\in\mathcal{N}(s_n[t])$, and receives reward $r_n[t]$.

GC-QAP uses an $\epsilon$-greedy policy to find the neighbors us
\begin{align}
a_n[t]= 
\begin{cases}
\displaystyle \arg\max\limits_{a\in \mathcal{A}(s_n[t])} Q_n\!\big(s_n[t],a\big), & \text{for~~~} 1-\epsilon,\\[1.2ex]
\text{Unif}\big(\mathcal{A}(s_n[t])\big), & \text{for~~} \epsilon.
\end{cases}
\label{eq:eps-greedy}
\end{align}
As $\zeta\in(0,1)$ is the discount factor and $\alpha\in(0,1]$ the step size. With next state $s_n'[t]$ and reward $r_n[t]$, the temporal-difference (td) target and update are
\begin{align}
y_n[t] 
&= r_n[t] \;+\; \zeta \max_{a'\in \mathcal{A}(s_n'[t])} Q_n\!\big(s_n'[t],a'\big),
\label{eq:td-target}\\
Q_n(s_n[t],a_n[t]) 
&\leftarrow (1-\alpha)\,Q_n(s_n[t],a_n[t]) \;+\; \alpha\, y_n[t].
\label{eq:q-update}
\end{align}
The GC-QAP formulation yields a discrete, low-dimensional MDP after graph condensation, so Q-learning is lightweight, stable, and sample-efficient. On the other hand, GC-QAP did not adopt the PPO/Deep Deterministic Policy Gradient (ddpg), as this would mainly add computational cost without clear benefit until GC-QAP flight dynamics are continuous, which is in our future effort.
GC-QAP restricts the actions by graph adjacency and motion limits by enforcing $\mathcal{A}(s) \;=\; \Big\{ a\in \mathcal{N}(s)\ \big|\ \|\mathbf{q}_n(a)-\mathbf{q}_n(s)\|_2 \le v_{\max}\Delta_t,\ 
h_{\min}\!\le\!h_n\!\le\!h_{\max} \Big\},$
so the $\max$ in \eqref{eq:td-target} and the policy in \eqref{eq:eps-greedy} are always taken over feasible moves.
\subsection{Proposed Quantum-inspired Condensed Nodes Computation}
Computations process a dense set of candidate ground nodes $
\mathcal{V} = \{ \mathbf{v}_1, \mathbf{v}_2, \dots, \mathbf{v}_{N_0}\}, 
\quad \mathbf{v}_i \in \mathbb{R}^2,$ where $N_0$ is the total number of initial deployment points. 
To make trajectory optimization tractable, this dense set is reduced to a smaller subset $
\mathcal{C} = \{ \mathbf{c}_1, \mathbf{c}_2, \dots, \mathbf{c}_N \}, 
\quad N \ll N_0,$ which helps to improve the coverage fidelity and connectivity structure. While conventional condensation approaches, such as $k$-means clustering, can be used as baselines~\cite{8833519}, we propose a QA-inspired condensation scheme. The condensation problem minimizes the clustering distortion cost $\mathcal{L}(\mathbf{c}_j) = 
\sum_{i=1}^{N_0} \min_{j \in \{1,\dots,N\}} \|\mathbf{v}_i - \mathbf{c}_j\|_2^2,$ which measures the fidelity of representing the original dense graph by $N$ centroids. Unlike other algorithms, QA-inspired condensation employs a stochastic annealing process that mimics quantum tunneling effects to escape local minima. At each iteration, candidate centroid positions are perturbed not only by small local updates but also by quantum jumps, which allow non-local moves to unexplored regions of the solution space. A candidate update $\Delta \mathcal{L}$ in cost is accepted with probability $P_{\text{accept}} = \exp(-\Delta \mathcal{L}/T),$ where $T$ is the annealing temperature controlling exploration. 
The temperature $T$ is gradually reduced according to a cooling schedule, thereby transitioning from explorative quantum jumps to fine-grained refinements. This hybrid exploration–exploitation strategy allows the QA-inspired scheme to more effectively search solution than classical annealing. The algorithm terminates when either a maximum number of iterations is reached or when $\mathcal{L}$ stabilizes within a tolerance window. The resulting condensed set $\mathcal{C}$ contains $N$ representative centroids that capture the dominant coverage structure 
of the original dense set $\mathcal{V}$ while significantly reducing 
the computational complexity of UAV trajectory optimization. 

\section{Proposed Graph Condensation (GC) via Quantum–Annealing Placement (QAP) for Trajectory Optimization}
\label{sec:proposal}
This paper proposes GC strategy via QAP optimization to reduce the UAV state space while enforcing priority-aware service. The condensed waypoint set $\mathcal{C}$ defines the discrete state–action space for the RL agent; at each time step, users are associated by instantaneous SINR, and uplink inter-cell interference is mitigated via power control. The detailed steps of the proposed algorithm is summarized in Algorithm \ref{alg:QA_RL}. 

As a baseline, GC with signal-to-noise ratio (snr)-based prioritization (GC-SNRP) is implemented to prune the waypoint set to the most promising UAV locations. An SNR proxy metric is computed from large-scale terms such as path loss, LoS probability, and received power, without modeling inter-cell interference with optional up-weighting for priority users. Candidates are then ranked and greedily selected under a spatial-diversity constraint, yielding a compact set of high-quality centroids. This shrinks the RL state/action space while preserving coverage, enabling more efficient trajectories, lower outage, and better priority protection than $k$-means condensation.

To benchmark performance, results also present a GC-k-means baseline, where candidate points are clustered and the resulting $|\mathcal{C}|$ centroids define feasible UAV placements. Each UAV runs independent Q-Learning (IQL) on this condensed graph. Even when user priorities are incorporated into the learning stage, GC-k-means remains fundamentally limited because the condensation step itself only captures average spatial density and ignores radio-aware factors such as SINR reliability and interference hot-spots. As a result, many selected centroids may lie in regions that are geometrically representative but suboptimal for coverage and reliability, leading to higher outage compared to the proposed GC-QAP scheme. By embedding SNR-based prioritization and quantum-inspired exploration directly in the condensation process, GC-QAP produces centroids that emphasize strong coverage for priority users and mitigate interference, enabling consistently better outage and rate performance under both priority-aware and non-priority settings.

Compared with GC-k-means and GC-SNRP, GC-QAP explores a broader basin of candidate subsets via temperature–controlled uphill moves, producing waypoint sets that better balance coverage, separation, and graph connectivity. The proposed approach significantly reduces the outage probability and improves spectral efficiency compared to conventional k-means and SNR condensation.

\begin{algorithm}[!t]
\scriptsize
\caption{Proposed GC-QAP for Multi-UAV Trajectory Optimization}
\label{alg:QA_RL}
\textbf{Input:} Candidate nodes $\mathcal{V}$; users $\mathcal{K}$ with weights $w_i$ and priority labels; UAVs $N$; SINR threshold $\gamma_{\sf th}$; noise power $\sigma^2$;\\
\textbf{Output:} Optimized UAV trajectories $\{\mathbf{q}_n[t]\}_{t=1}^T$.

\medskip
\textbf{Step 1: Centroid condensation (QA)}\\
Initialize centroid set $\mathcal{C}$ from $\mathcal{V}$, set $T=T_0$.\\
repeat\\
\quad Propose a quantum jump: perturb one centroid in $\mathcal{C}$.\\
\quad Compute clustering distortion\\
\quad Accept new set with probability $\min\{1,\exp(-(J_{\text{new}}-J_{\text{old}})/T)\}$.\\
\quad Update $T \leftarrow \rho T$.\\
until $T<T_{\min}$.\\
Return condensed set $\mathcal{C}$ and adjacency $\mathcal{N}(\cdot)$.

\medskip
\textbf{Step 2: RL-based trajectory optimization}\\
for episode $e=1:E$ do\\
\quad Initialize UAVs at states $s_n[1]\in\mathcal{C}$.\\
\quad for time-step $t=1:\tau$ do\\
\quad\quad For each UAV $n$, select action $a_n[t]$ by $\epsilon$-greedy on $Q_n(s_n[t],a)$.\\
\quad\quad Move UAV to $\mathbf{q}_n[t{+}1]\in\mathcal{N}(s_n[t])$, obeying velocity and altitude limits.\\
\quad\quad Apply open-loop power control for each user $i$ as computed in system model, and compute SINR using  $\gamma_{i}[t] = \frac{p^{\sf tx}_i[t]\; g_{i,n}[t]}{\sigma^2 + I_{n,i}[t]}$\\ 
\quad\quad Associate user $i$ with UAV based on max. received power\\
\quad\quad Compute UAV reward using \eqref{eq:reward2} \\
\quad\quad Update Q-table
\begin{align}
Q_n(s_n[t],a_n[t]) \leftarrow & (1-\alpha_Q)Q_n(s_n[t],a_n[t]) \nonumber\\
& + \alpha_Q \Big( r_n[t] + \gamma_Q \max_{a'} Q_n(s'_n[t],a') \Big). \nonumber
\end{align}
\quad end for\\
end for\\
Return optimized UAV trajectories $\{\mathbf{q}_n[t]\}$.
\end{algorithm}

\begin{figure*}[!t]
\centering
\subfloat[]{%
    \includegraphics[width=0.24\textwidth]{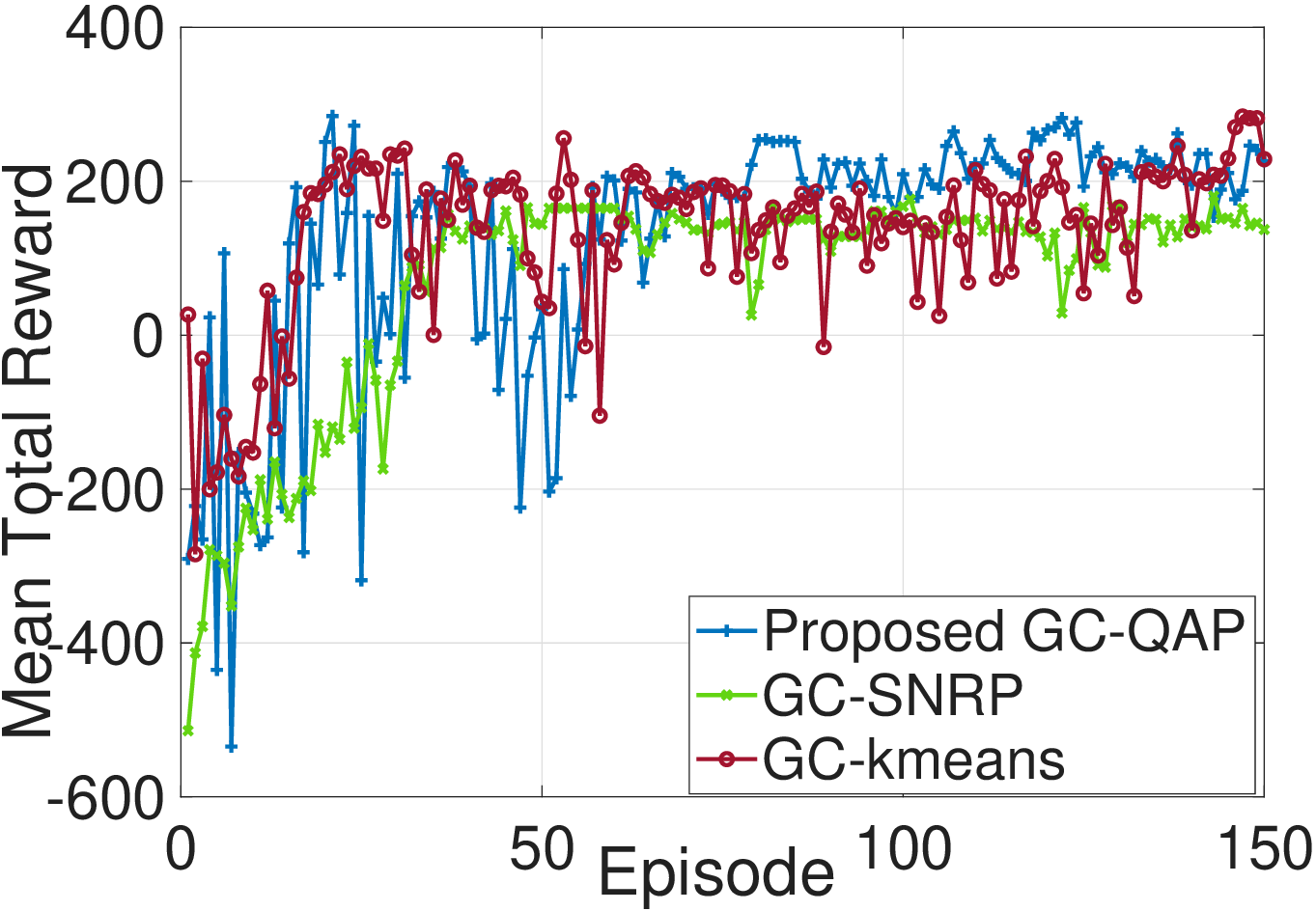}%
    \label{fig:reward}
} \hfil
\subfloat[]{%
    \includegraphics[width=0.24\textwidth]{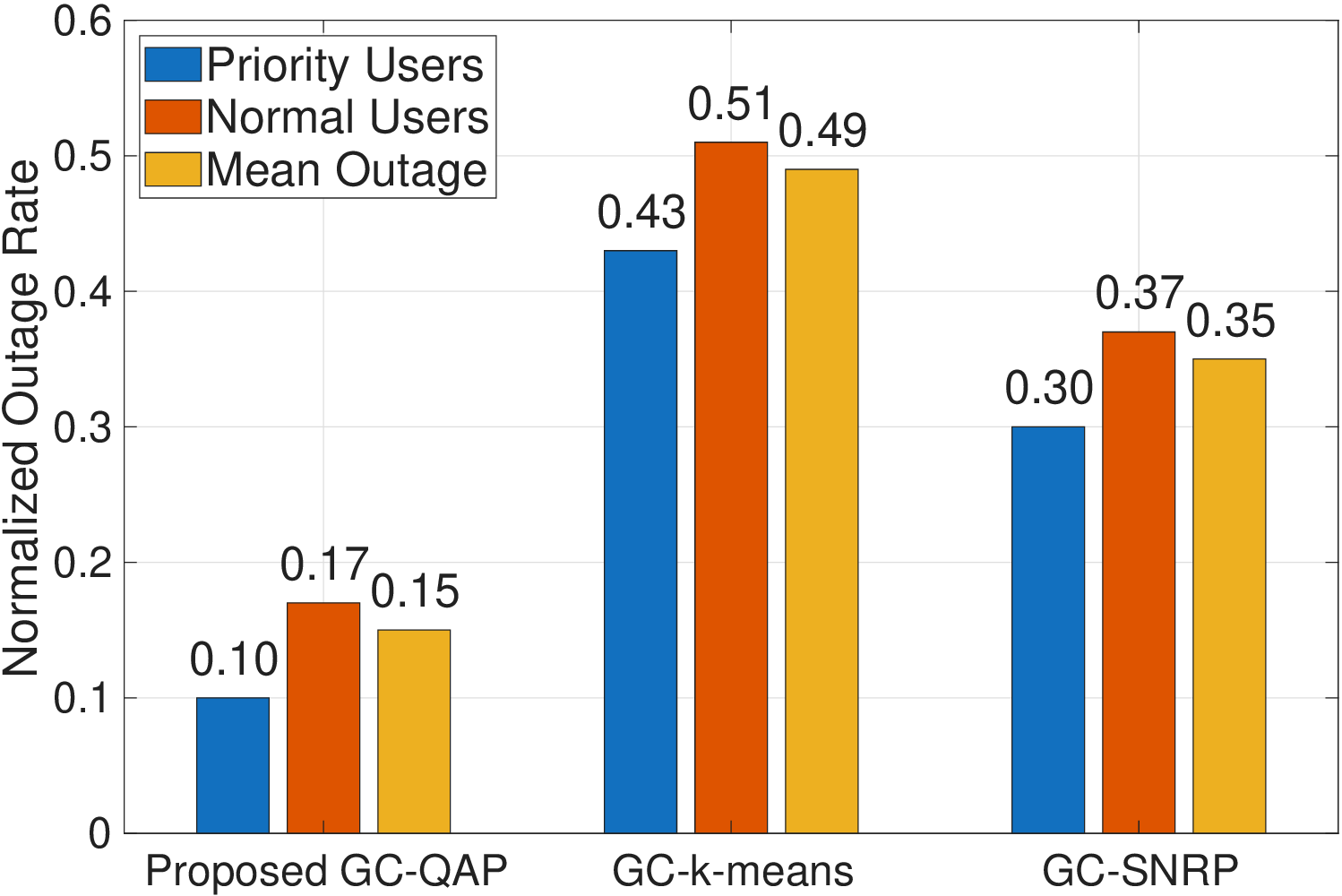}%
    \label{fig:outage}
} \hfil
\subfloat[]{%
    \includegraphics[width=0.24\textwidth]{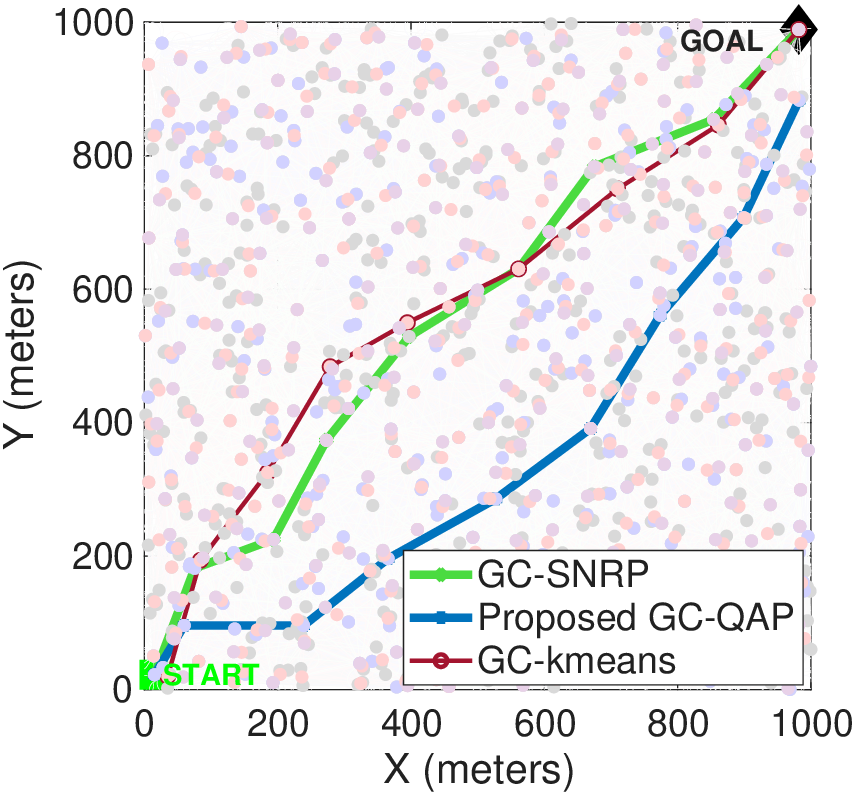}%
    \label{fig:traj}
}
\hfil
\subfloat[]{%
    \includegraphics[width=0.24\textwidth]{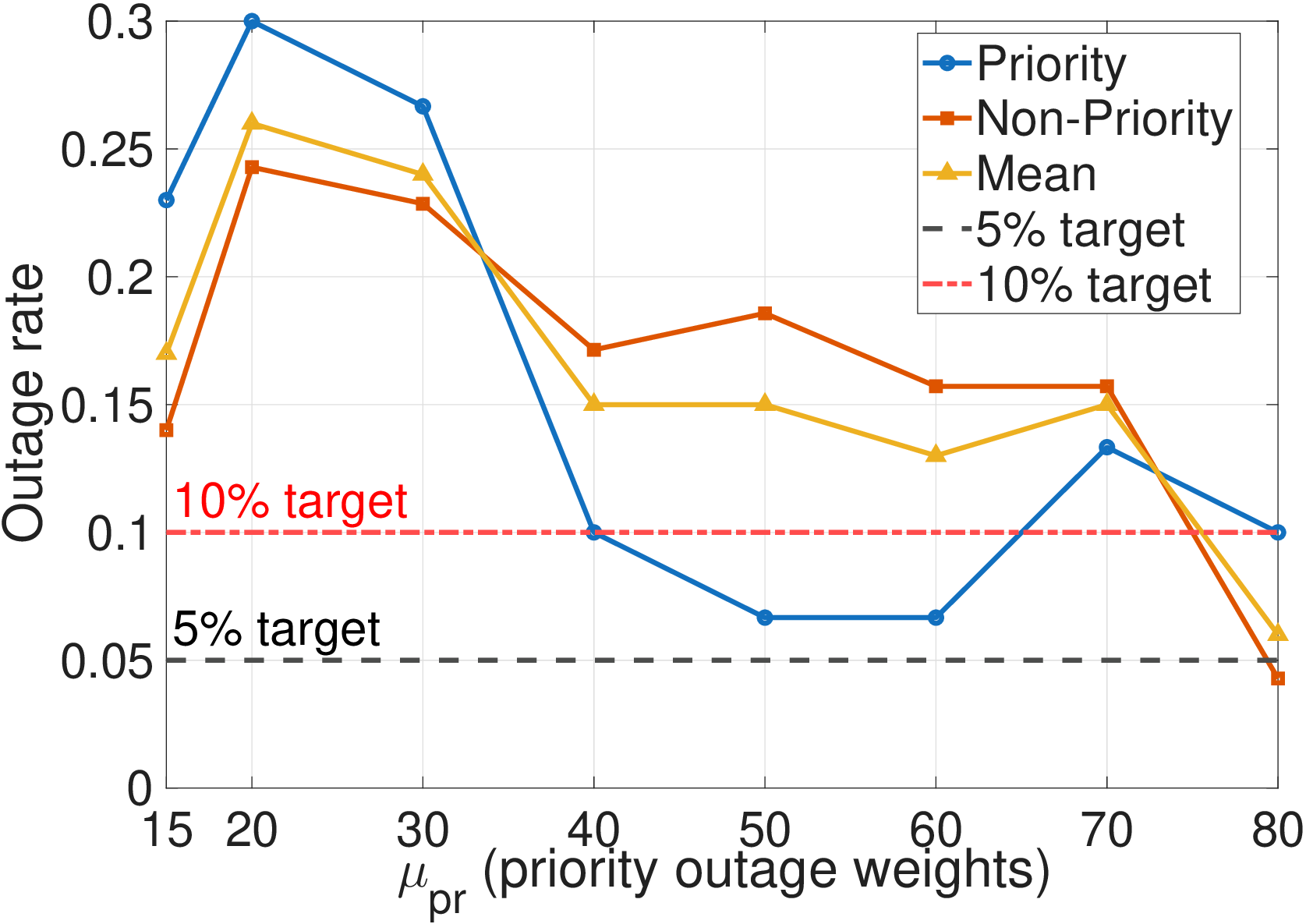}%
    }
\caption{Performance evaluation under different schemes:
(a) reward vs. episode, (b) User outage comparison, and (c) Optimized UAV trajectory, (d) outage vs. priority-penalty weight $\mu_{\rm pr}$.}
\label{fig2}
\vspace{-0.15in}
\end{figure*}

\begin{figure}[!b]
\centering
\centering
\subfloat[]
{\includegraphics[width=0.46\columnwidth]{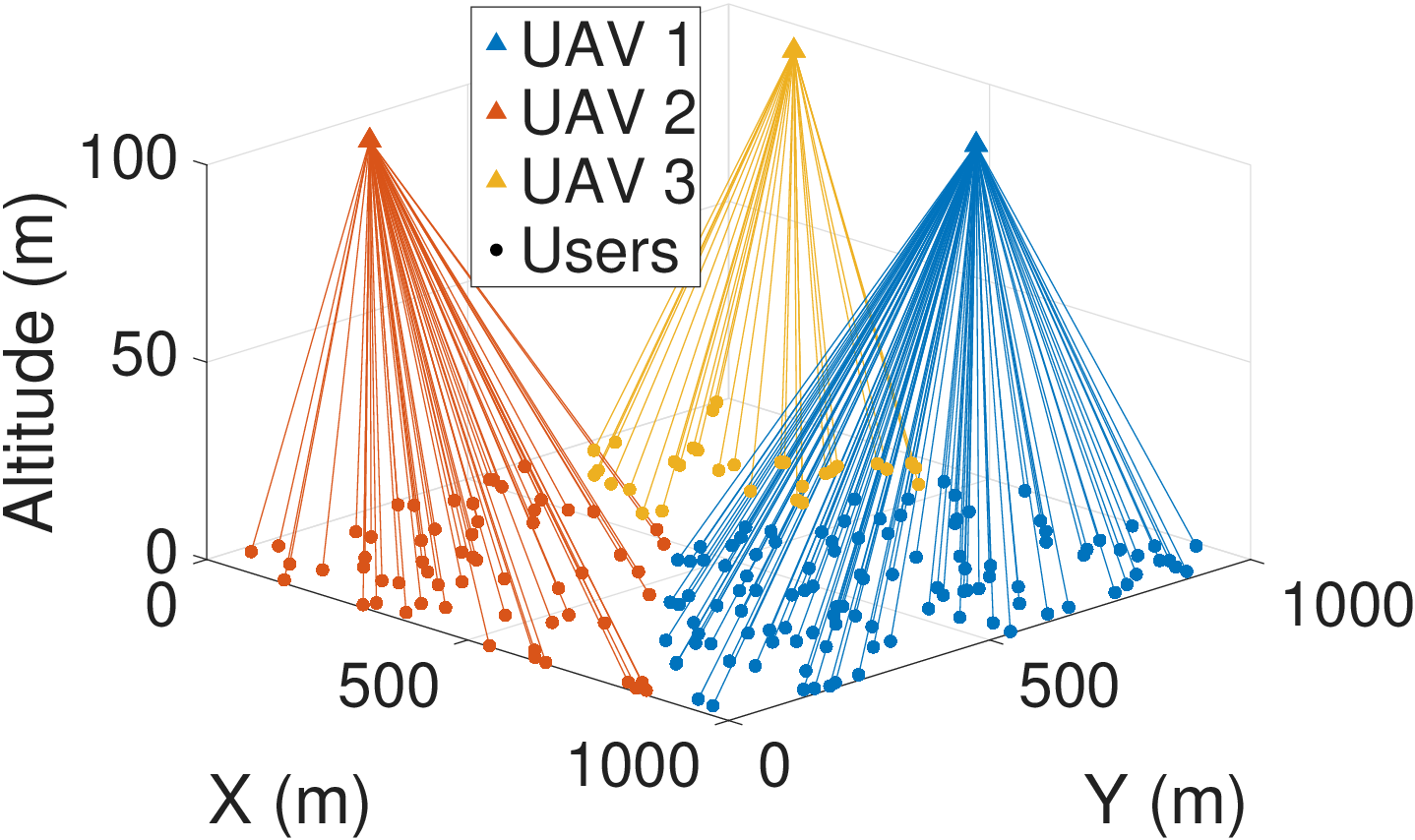}}%
\label{fd}
\hfil
\subfloat[]
{\includegraphics[width=0.46\columnwidth]{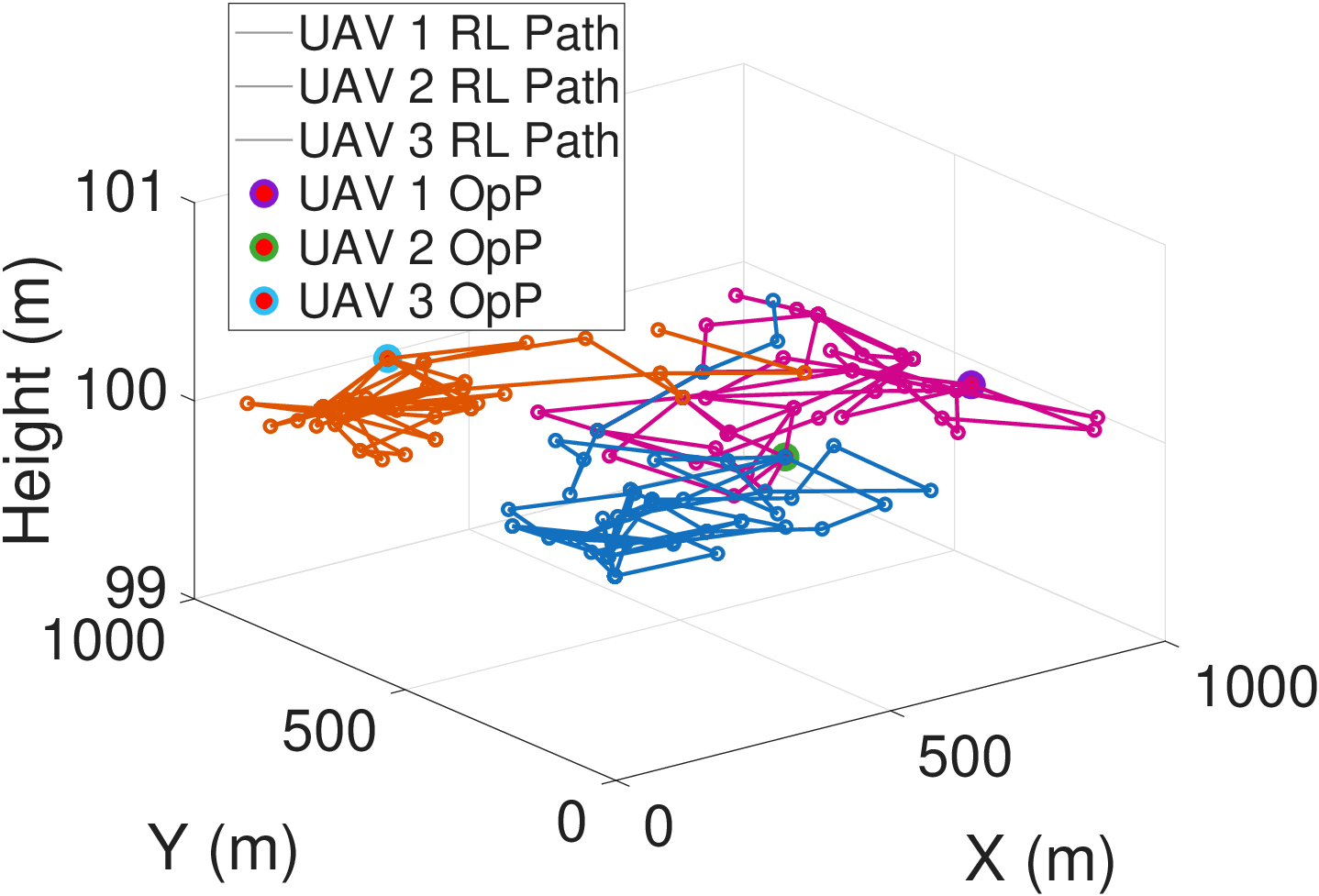}}%
\label{fd}
\hfil
\caption{UAV trajectory optimization phase for the proposed GC-QAP scheme, (a) 3D association, (b)  trajectory optimization.}
\label{fig}
\vspace{-0.25in}
\end{figure}

\section{Simulation Results}
To evaluate GC-QAP performance, a low-altitude multi-UAV communication network is simulated in MATLAB environment by deploying over a $1000 \times 1000$~m$^2$ region. The proposed GC-QAP uses a quantum-inspired annealing algorithm implemented on classical hardware, emulating quantum tunneling and global search to escape local minima.

A total of $N=3$ UAVs were deployed at a fixed altitude of $h=100$~m, serving $K=100$ ground users randomly distributed within the coverage area, where $20\%$ of users were designated as priority users. The candidate waypoint graph is condensed to $33$ centroids using quantum annealing, k-means, or SNR-based clustering for comparison. The wireless channel incorporated both LoS and NLoS components with parameters $b_1=0.1$, $b_2=1$, $\xi=5$, path-loss exponent $\alpha=2.0$, and additional losses $\kappa_{\sf LoS}=1$~dB, $\kappa_{\sf NLoS}=20$~dB. The carrier frequency is $f_c=2$~GHz and the noise power is set to $-90$~dBm. Each user’s transmit power is limited by $P_{\max}=23$~dBm with open-loop power control. The system bandwidth is $B=1$~MHz, and the SINR threshold is fixed at $\gamma_{\sf th}=5$~dB. For quantum-annealing clustering, adopted values are $T_{0}=100$, $T_{\min}=10^{-3}$, cooling rate $\rho=0.95$, \; $I_{\max}=1000$, with priority weightage selected as $\mu_{\mathrm{pr}} =40$.
RL model was trained over $400$ episodes with $100$ time-steps per episode, where UAV mobility was constrained by adjacency and maximum speed limits $v_{\max}\Delta_t$. 

Fig.~\ref{fig2}(a) present the reward that increases with the number of priority-weighted users meeting the SINR threshold and decreases with outage penalties. Hence, larger values indicate better reliability. GC-QAP converges rapidly and stabilizes around 200–250, while GC-SNRP reaches near 120–160 and GC-$k$-means shows slower, higher-variance convergence. The gap reflects GC-QAP’s ability to steer UAVs toward radio-favorable centroids, yielding fewer outages under the same training budget. For the proposed GC-QAP graph, the transitions are smoother and more Markovian which leads to more stable learning and faster convergence.

Moreover, Fig. \ref{fig2}(b) presented the normalized outage rates for priority users, non-priority users, and the overall mean. The proposed GC-QAP achieves the mean outage of $0.15$, significantly outperforming GC-k-means ($0.49$) and GC-SNRP ($0.35$). In comparison, GC-QAP reduces the mean outage by $69.4\%$ compared to GC-k-means and by $57.1\%$ compared to GC-SNRP.
GC-QAP also lowers the priority-user outage to $0.1$, compared to $0.43$ for GC-k-means and $0.30$ for GC-SNRP, yielding 46.5\% and 23.3\% reductions, respectively. For non-priority users, GC-QAP achieves $0.14$ compared to $0.51$ for GC-k-means and $0.37$ for GC-SNRP, i.e., 76.7\% and 66.7\% reductions. Thus, the proposed method consistently lowers outages for both user groups.
These improvements are because QA condensation produces graphs whose connectivity aligns with user density and channel reliability, the reward shaping penalizes priority outages strongly, and SINR with inter-cell interference guides UAVs to avoid interference.
Finally, Fig.~\ref{fig2}(c) illustrates the trajectory of a single UAV under different condensation schemes. While GC-k-means and GC-SNRP reduce the graph size, they either neglect channel quality or generate interference-prone zig-zag paths. Fig.~\ref{fig2}(d) shows outage versus the priority penalty $\mu_{\mathrm{pr}}$. Increasing $\mu_{\mathrm{pr}}$ from $15$ to $60$ reduces priority-user outage from $0.30$ to $0.07$, results in crossing the $10\%$ target and entering the $5$–$10\%$ band, while non-priority outage stays $\le 0.18$ and can drop to $0.045$ at $\mu_{\mathrm{pr}}{=}80$; the mean outage is around 0.06. Hence, GC-QAP can meet fixed reliability targets for priority users by tuning a single weight. GC-QAP consider best-effort enhanced mobile broadband service, for which a $5$–$10\%$ outage target is commonly used.

In contrast, the proposed GC-QAP steers the UAV through radio-aware centroids aligned with user clusters and LoS-favorable corridors, achieving higher SINR and lower outage, as further supported by the 3D association and trajectory optimization results in Fig.~\ref{fig}. 
The computational complexity of the considered schemes are discussed and validated during real-time simulation. As all schemes employ the same RL backbone, the primary complexity difference arises from the graph condensation stage. GC-k-means \cite{1017616} and GC-SNRP scale as $\mathcal{O}(I_{\text{km}} N C)$ and $\mathcal{O}(I_{\text{snr}} N K)$, respectively, while the proposed GC-QAP performs probabilistic tunneling over $C$ centroids only, achieving $\mathcal{O}(I_{\text{QA}} C)$ complexity independent of $N$. Empirically, the QA condensation completed in  around 0.21~s, compared to 1.14~s for GC-k-means  and 2.06~s for GC-SNRP.
Although, all schemes exhibit almost similar RL training time, confirming that GC-QAP’s improvement emerged from its faster condensation process. Overall, the proposed GC-QAP framework achieves a practical balance between computational efficiency and optimization accuracy, making it suitable for real-time UAV trajectory planning in low-altitude networks.

\section{Conclusion}
The proposed GC-QAP framework achieved reliable and efficient UAV trajectory optimization by integrating quantum-inspired condensation with RL. It delivers up to 77\% lower priority-user outage and 69\% lower mean outage compared to GC-k-means and GC-SNRP, while maintaining geometrically efficient paths. By reshaping the search space into radio-aware centroids, GC-QAP balances coverage, interference management, and scalability, making it a practical solution for UAV-assisted low-altitude networks. The proposed GC-QAP framework employs a \emph{quantum-inspired annealing} algorithm implemented on classical hardware as no quantum hardware acceleration is assumed. The approach mimics key quantum principles—probabilistic tunneling and global sampling—to escape local minima while retaining full classical reproducibility. Future work will deploy the annealing kernel on real quantum annealers, e.g., D-Wave, to evaluate practical quantum real-time computation.

\ifCLASSOPTIONcaptionsoff
  \newpage
\fi

\bibliographystyle{IEEEtran}
\bibliography{Biblio}
\end{document}